# Going Beyond the Debye Length:
# Overcoming Charge Screening Limitations in Next-Generation Bioelectronic Sensors

**Authors:** Vladimir Kesler[1], Boris Murmann[1], H. Tom Soh [1,2,3]

[1] *Department of Electrical Engineering, Stanford University, Stanford, CA 94305, USA.*
[2] *Department of Radiology, Stanford University, Stanford, CA 94305, USA.*
[3] *Chan Zuckerberg Biohub, San Francisco, CA 94158, USA.*

**Abstract:**
Electronic biosensors are a natural fit for field-deployable diagnostic devices, because they can be miniaturized, mass produced, and integrated with circuitry. Unfortunately, progress in the development of such platforms has been hindered by the fact that mobile ions present in biological samples screen charges from the target molecule, greatly reducing sensor sensitivity. Under physiological conditions, the thickness of the resulting electric double layer is less than 1 nm, and it has generally been assumed that electronic detection beyond this distance is virtually impossible. However, a few recently-described sensor design strategies seem to defy this conventional wisdom, exploiting the physics of electrical double layers in ways that traditional models do not capture. In the first strategy, charge screening is decreased by constraining the space in which double layers can form. The second strategy uses external stimuli to prevent double layers from reaching equilibrium, thereby effectively reducing charge screening. The goal of this article is to describe these relatively new concepts, and to offer theoretical insights into mechanisms that may enable electronic biosensing beyond the double-layer. If these concepts can be further developed and translated into practical electronic biosensors, we foresee exciting opportunities for the next generation of diagnostic technologies.

**The tyranny of double layer screening**

Over the past decade, there has been exciting progress in leveraging electronics for biomolecular detection. This movement has produced many innovations in medical technologies, such as semiconductor-based DNA sequencing[1] and continuous glucose monitors[2,3]. Electronic biosensors are a natural fit for point-of-care diagnostic technologies due to their potential for miniaturization, low cost of manufacturing, and integration with advanced electronics. In most electronic biosensors, electrodes typically serve as interfaces between electrical and biological signals. By modifying an electrode with affinity reagents, such as antibodies or aptamers, researchers can design interfaces that specifically bind to biomolecular analytes of interest. However, the sensitivity of electronic biosensors is limited by screening of electric fields by mobile ions[4,5]. All biological samples contain high concentrations of such ions, and as a consequence, this screening effect can greatly attenuate biosensor signals. This fundamental physical effect, which manifests as the electric "double layer", prevents many electronic detection platforms from becoming broadly useful. Researchers looking to model screening effects at the electrode-electrolyte interface commonly rely on a simple capacitive model based on the linearized Poisson-Boltzmann formulation for decaying potentials in the presence of mobile charges[6,7] (Box 1). The Debye length arises from this model, providing a helpful yardstick for the spatial scale of such screening. The Debye length under physiological conditions is less than 1 nm; in contrast, the length of an antibody is on the order of 10-15 nanometers, and a 30-base aptamer is up to ~10 nm[8]. This intrinsic mismatch in dimensions between charge screening and size of biomolecules poses a fundamental challenge for biosensors based on these reagents (Fig. 1).

Despite the challenge posed by double layer screening, researchers have implemented innovative electronic biosensors for decades. There are three common ways to measure binding as an electrical signal: i) measuring a change in potential at the interface due to the movement of charges[1,9], ii) measuring a change in impedance due to the presence of the biomolecule at the interface[10,11], and iii) measuring an electrochemical signal generated by a reporter molecule that can be interrogated with electric fields[12,13]. However, progress is slowing down because most ideas for mitigating screening, in all three types of biosensors, are being exhausted.

Does this mean that the field of electronic biosensing is nearing a dead end—a tantalizing prospect doomed to never live up to its full potential? Fortunately, we believe the answer to this question is a resounding no, and that our simple models for understanding charge screening – including the Poisson-Boltzmann model – have merely reached the limits of their utility. We further believe there are important physical phenomena that we are only beginning to uncover that will enable biomolecular detection far beyond the Debye length, yielding a more sophisticated next generation of electronic biosensing platforms.

Recent efforts offer interesting perspectives on overcoming conventional charge screening limitations through two concepts: the Debye volume and non-equilibrium measurement techniques. Researchers have already begun applying these concepts to develop improved biosensors that achieve a level of performance that Poisson-Boltzmann modeling cannot accurately capture. In this Perspective, we outline these strategies and summarize the current state of the art. By calling attention to these still-underutilized strategies for overcoming double layer screening, we hope to motivate future innovation in electronic biomolecular detection.

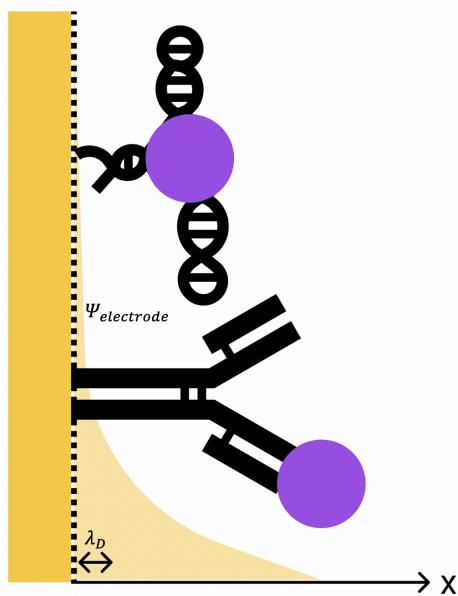

**Figure 1: Electrical double layers form in all electronic biosensors.** The Debye length ($\lambda_D$) is the characteristic distance of potential decay in electrolyte solutions and is much shorter than common molecular receptors. The potential decay in the electrode's double layer ($\Psi_{electrode}$) is shown in comparison with two example receptors: an aptamer (top) and antibody (bottom). Due to this mismatch in dimensions, it is challenging to generate an electronic signal based on binding with target molecules.

## The Debye volume: limiting screening through surface engineering

Easily overlooked in discussions of charge screening is the fact that double layers take up space. In fact, an electric double layer behaves more like a diffusion capacitor than a simple parallel plate capacitor: it extends outward in space from the charge it screens. By acknowledging this fact, researchers gain an additional control knob for designing the next generation of electronic biosensors.

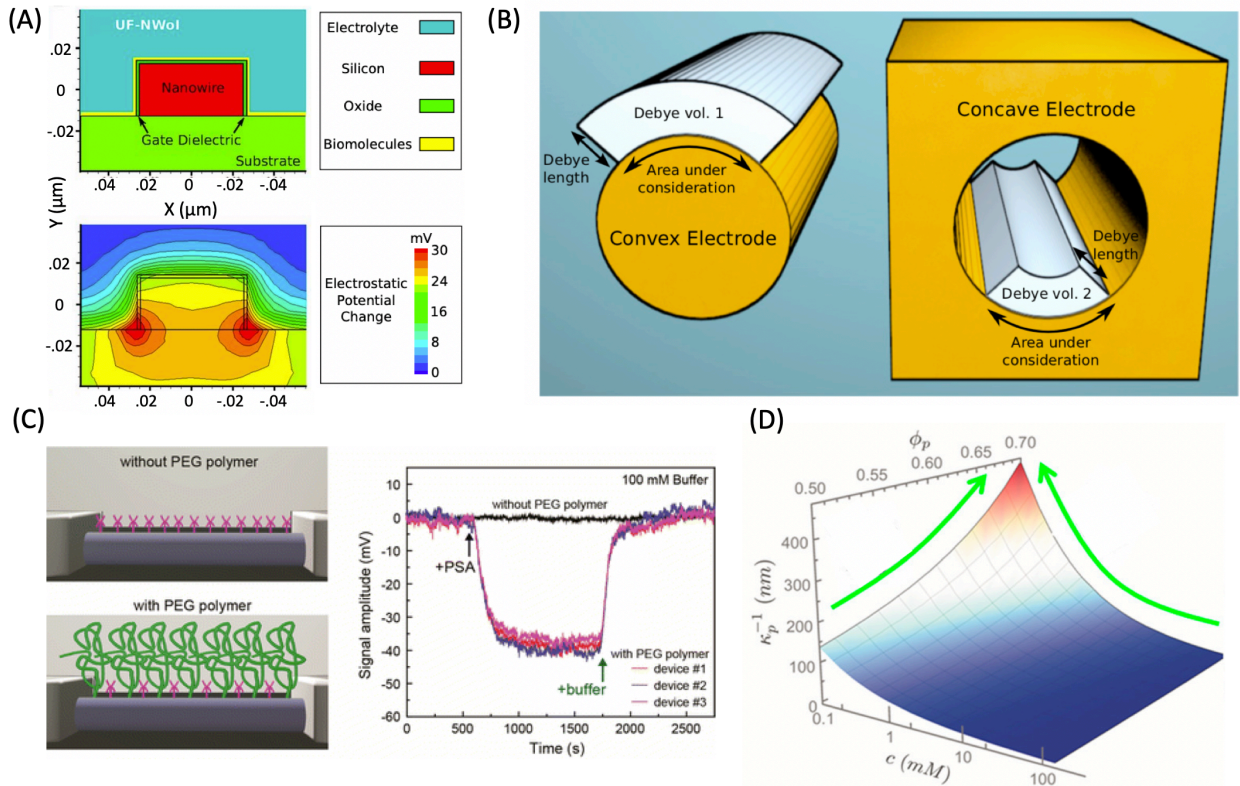

**Figure 2: Debye volume-based approaches to reduce screening.** (A) Simulations of a nanowire field-effect transistor (NW-FET) biosensor reveal concave corner regions where the screening effect is reduced (red)[14] (B) The Debye volume-to-surface area ratio provides a conceptual framework for evaluating screening near electrode-electrolyte interfaces. A higher ratio of volume to surface area corresponds to stronger screening for the convex electrode (left), while the converse is true for the concave electrode (right).[14] (C) An alternative way to occupy Debye volume is to coat the sensing surface with large polymers, auch as PEG. Binding sites are shown in red; surface immobilized PEG molecules are shown in green (left). Immobilizing these polymers on sensing surfaces dramatically improves sensitivity (right)[15]. Specific detection can be achieved by co-immobilizing these polymers with aptamers[16] or antibodies[17]. (D) Screening can be tuned by altering ionic strength of the solution and the polymer volume fraction of the surface coating ($\phi_p$). The effective screening length on polymer coated surfaces ($\lambda_p = \kappa_p^{-1}$) increases with decreasing ionic concentration ($c$) but also with increasing polymer volume fraction[18].

By limiting the volume available for ions to form double layers, it is possible to design surfaces where the double layer extends far beyond the Debye length. In 2014, Shoorideh and Chui conducted a series of simulations of various sensor geometries to explore whether nanowire field-effect transistor (NW-FET) biosensors were more sensitive than planar ones due to their higher surface-to-volume ratio[19] (Fig. 2A). Simulations revealed that this was not the case, but the authors also showed that a nanowire structure placed on top of a flat substrate inadvertently introduces concave corners where the screening effect was reduced. To help conceptualize this effect, the authors introduced the concept of Debye volume (Fig. 2B), which is the volume encompassed by a surface drawn one Debye length away from the electrode, normal to the surface. The ratio between this volume and the surface area of the electrode can help visualize the change in sensitivity. For a convex surface, the Debye volume-to-surface area ratio is high and there is more space around the electrode for ions to approach and screen surface charges. In contrast, the ratio is lower for concave electrode surfaces, and the reduced volume introduces energetic constraints that reduce screening. This observation can also be examined mathematically, albeit with more effort, by examining the Poisson-Boltzmann formulations for the diffuse layer capacitance of various electrode geometries.

Although the concept of Debye volume has not been broadly established, it is a useful tool for analyzing complex biosensing interfaces that makes it easier to understand why certain sensor geometries are more sensitive than others. Nanogap and nanopore electrodes are good examples of interfaces where double layers crowd one another, reducing screening[20,21]. These structures come with their own challenges, however; they are more difficult to manufacture than planar electrodes, and their small dimensions limit the scope of targets they can detect and make them susceptible to biofouling. However, there are other ways to restrict Debye volume at the electrode interface. For example, the work by Gao *et al.* were the first to report a description of a surface coating that decreases screening on FET-based biosensors[15,16]. After coating the surface of their electrodes with large polyethylene glycol (PEG) molecules (Fig. 2C), they were able to detect prostate-specific antigen (PSA) in physiological ionic strength buffers[15]. They achieved improved specificity by co-immobilizing a PSA-specific aptamer onto the electrode alongside the PEG coating[16]. Without the PEG coating, the non-specific devices could not detect PSA and the sensitivity of the aptamer-immobilized devices was reduced five-fold. Gutierrez-Sanz *et al.*

extended this approach to detect thyroid-stimulating hormone with antibodies in undiluted serum,[17,22] reporting a three-fold improvement in sensitivity after adding large PEG molecules to the electrode surface. Song *et al.* also employed a PEG coating to detect glial fibrillary acidic protein[23]. Their study included an investigation of optimal molecular weight of the PEG molecules and showed a trend towards higher sensitivity with higher molecular weight PEG. However, this sensitivity comes at a cost: binding kinetics between the sensor and the analyte are observably slower when using the PEG coating, as biomolecules must diffuse through the dense coating to be detected. Fortunately, Gao *et al.* have also shown that this penalty merely delays the final signal from ~3 minutes to ~15 minutes[16], which is sufficiently rapid for point-of-care applications. Gao *et al.* attribute the improvement in sensitivity to the change in interfacial capacitance due to the partially hydrated PEG coating[15,16]. This change in sensitivity has also been explained in terms of Donnan potential theory, which describes screening through dense, ion-permeable layers[24]. Unfortunately, neither of these explanations are fully satisfying – the former relies on an overly simple capacitive model of the interface, while the latter relies on impractically complicated math.

A more intuitive solution would be to apply the concept of Debye volume to these biosensing interfaces. A charged molecule surrounded by a dense, partially hydrated layer of PEG has limited volume for ions to approach and give rise to screening. As a result, the fields emanating from charges within the PEG layer persist farther away, exceeding traditional predictions of Debye length. This interpretation is supported by the work of Piccinini *et al.*, who explored the increase in sensitivity of graphene FET biosensors when coated with polyelectrolyte multilayers (PEM)[18]. They experimentally showed detection beyond the Debye length of polyelectrolyte layers of opposite charges as they were assembled on top of the FET surface. They also developed a thermodynamic theory that attributes this increase in sensitivity to the entropic cost of confining ions inside the PEM, and used this model to predict the Debye length inside such layers with varying polymer volume fractions (Fig. 2D). Their model shows that higher volume fractions lead to longer Debye lengths, and their PEMs — with a polymer volume fraction of .68 compared to .2 for PEG — can increase the screening length by an order of magnitude. Although they do not experimentally validate the detection performance, their analysis supports the Debye volume model of screening behavior.

**Disrupting the double layer through electronic perturbation**

Poisson-Boltzmann models describe screening in terms of the charge being buried beneath counter-ions. This is similar to looking for underwater seashells beneath a layer of sand. At equilibrium, the shells are buried and thus invisible; but when we kick the sand, they are visible until the sand settles. Much like the sand in this analogy, ions require some time in order to form double layers and screen potentials. Because ions have non-negligible size, they have finite diffusivity and mobility. This sets a lower bound on the time necessary for them to adjust to a change in potential—essentially, kicking the 'sand'—this relaxation time is referred to as the "Debye time"[25]. Additionally, the Poisson-Boltzmann case does not consider electric double layers in the presence of external ionic concentration gradients and fluid flow. For all of these cases, the Poisson equation must be coupled to the Nernst-Planck equation, which itself is an extension of Fick's law for the case of diffusion under the influence of electrostatic forces[26]. The coupled Poisson-Nernst-Planck (PNP) system of equations is analytically more complex, but allows for the exploration of circumstances in which the electronic environment surrounding electrodes and analytes is modulated by externally applied forces.

(A)
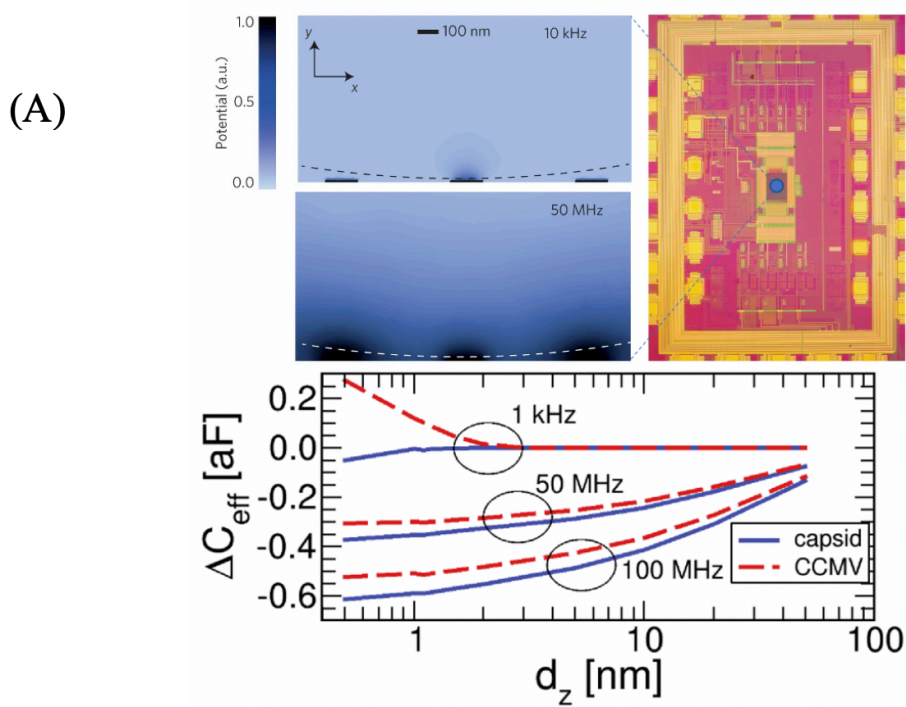

(B)

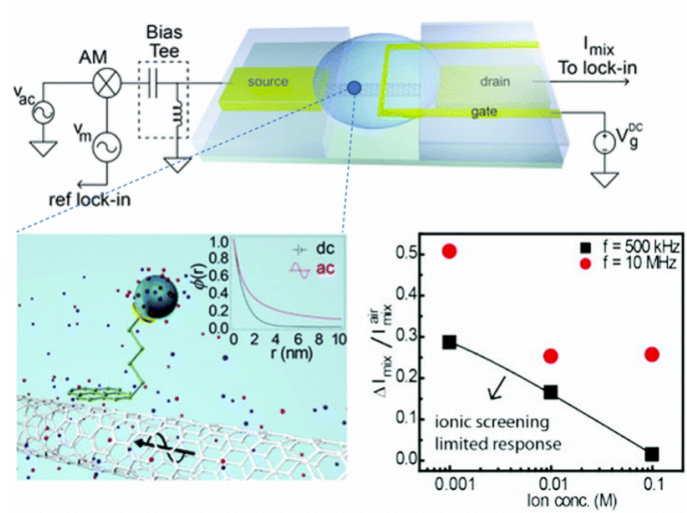

(C)

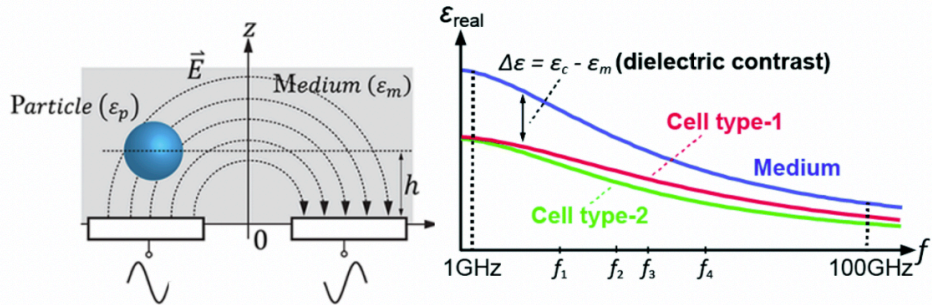

(D)

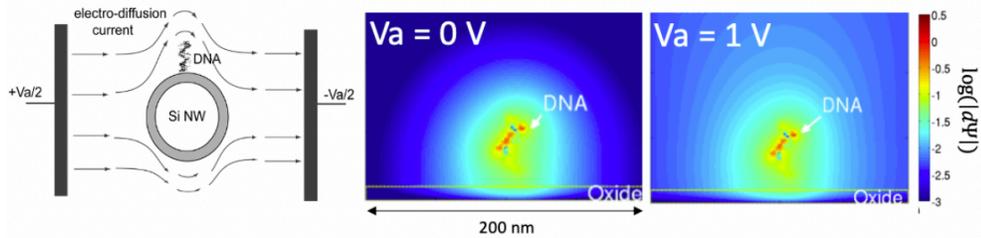

**Figure 3: Disrupting the electric double layer by forcing non-equilibrium screening** (A) A microsphere deposited on array of nanoelectrodes (top), which are measured as capacitors by an integrated circuit. At low operating frequencies (top left), the sphere can only be detected as a change in capacitance where it contacts the array, while the other pixels are screened close to the electrode. At high frequencies (bottom left), ion screening is perturbed, and the electrodes detect the microsphere[27]. When the same system is applied to cowpea chlorotic mottle virus (CCMV)(bottom), simulations show the capacitive response of the virus ($\Delta C_{eff}$) varies with distance from the electrode interface ($d_z$) and the frequency of the applied waveform (circles). The additional charge in the full CCMV particle relative to the empty capsid dominates the capacitive response in simulations at low frequencies. At high frequencies, response was governed by volume of the virus, which is identical to that of the capsid[28]. (B) A heterodyne biosensor operates by applying a high frequency waveform to the source of a nanoelectronic transistor through a bias tee circuit (top). This waveform stimulates biomolecular dipoles to oscillate, generating currents in the nanotube for detection (bottom-left) When the applied waveforms are at sufficiently high frequencies, the sensitivity

($\Delta I_{mix}/\Delta I_{mix}^{air}$) improves across physiologically relevant ionic concentrations (bottom-right)[29]. (C) A flow cytometry platform based on dielectric spectroscopy. When a particle passes through the electric field, it changes the dielectric constant of the solution (left). Different types of cells exhibit different dielectric constants ($\varepsilon_{real}$), which are measured across frequency to generate spectra (right)[30]. (D) Electro-diffusion current applied across a sensing interface weakens screening. A functionalized nanowire is placed between two electrodes that are used to generate electro-diffusion current when a potential (Va) is applied (left). When no potential is applied, no electro-diffusion is generated, and the signal caused by the charged DNA ($\log(|d\Psi|)$) decays quickly in normal double layers (center). When a potential is applied, the electro-diffusion current destabilizes the double layer, and electric fields persist farther into the solution (right) [31].

Impedance spectroscopy is a technique that exploits dynamic double layers[6]. Typically, a time-varying voltage waveform is applied to the electrochemical cell to generate an electrical current. The magnitude and phase of the electrical current are measured and then fit to discrete circuit elements to describe the impedance of the cell. In most biosensing applications, the frequency of the waveforms is low enough (<1 MHz) that ions have sufficient time to screen normally, but as the frequency of the applied waveform goes up, they fail to settle back into full equilibrium. Widdershoven *et al.* demonstrated that measuring double layer capacitive sensors via impedance spectroscopy at high frequencies allows them to see beyond the Debye length (Fig. 3A)[27,32–35]. They integrated a nanoelectrode array with readout circuitry to detect polymeric microspheres and live cells. When immersed in electrolyte solution, the nanoelectrodes form nanocapacitors by virtue of the electric double layers that form at the metal-electrolyte interface. The researchers then deposited microspheres on top of their array and analyzed them at various frequencies. At low frequencies, these beads are subject to equilibrium screening and can only be seen by pixels that are particularly close to where the sphere contacts the surface of the chip (Fig. 3A, top left). As the measurement frequency increases (up to 50 MHz), adjacent pixels can also detect the sphere because screening is reduced and the electric field from the electrode extends beyond the Debye length and couples from the solution into the microsphere (Fig. 3A, bottom left). Using this method, the researchers were able to discern beads of different dielectric properties and sizes as well as image different kinds of cells moving across the array. The nanocapacitors are measured by applying known voltages across them, and then unloading the charge into a transimpedance amplifier. This technique is analogous to common sampling circuits in integrated circuits. The innovation lies in the operation of the circuit – the rate of sampling can be varied up to tens of MHz, allowing exploration of screening effects across that range of frequencies. The authors also developed a robust simulation of this platform based on the PNP system of equations[36–39]. Using this system and a simple circuit model, they were able to analyze the change of impedance across frequencies[27]. Additionally, they simulated the response of such a system to virus particles, comparing the capacitance signal from a full cowpea chlorotic mottle virus with just the capsid from the same virus, which contained fewer charges[28]. This yielded the insight that at low frequencies, the signal is more sensitive to analyte charge, while at high frequencies, the signal is more sensitive to analyte volume. This makes intuitive sense: at high frequencies, a greater portion of the virus falls within the screening of the electrode potential, modulating the dielectric constant at that interface. At lower frequencies, the field does not penetrate as far, so whatever analyte charges are detectable under those conditions remain at the interface, fixed and unresponsive to the applied voltages.

Kulkarni *et al.* used high frequency stimuli to overcome Debye screening effects in a different way with their nanoelectronic heterodyne sensors (Fig. 3B)[29,40–43]. Here, a single-walled carbon nanotube-based transistor was used to detect streptavidin binding to biotin[29]. Unlike in a traditional FET-based biosensor, a high-frequency carrier (up to 30 MHz) is mixed with a lower frequency modulation signal and injected through the source terminal of the transistor. This signal causes any dipoles at the gate terminal to oscillate, generating a detection signal. To simplify their readout, they leveraged the transistor's non-linearity, which generates a low-frequency mixing current that reflects the magnitude of the dipoles' signal. In this way, their system gets the best of both worlds: simple low-frequency readout instrumentation and high-frequency screening behavior.

At still higher frequencies (in the GHz range), the double layer never forms, and the electric field extends all the way through the electrochemical cell. This makes the cell look like one capacitor, with the electrolyte as the dielectric material. This approach, known as dielectric spectroscopy[44,45], can be very sensitive and has been used to differentiate open and closed DNA-based molecular tweezers in buffer[46]. It is challenging to use this technique for specific biomolecular detection in biological samples, however, because any molecular changes in the complex sample can alter signals. But dielectric spectroscopy has proven suitable for other biomedical applications. For example, Chien *et al.* designed a high-frequency electronic flow cytometry readout (Fig. 3C)[30,47]. In their system, the capacitance between two electrodes modulates the resonant frequency of an oscillator. By operating at GHz frequencies, far beyond the frequency at which electric double layers can form, the fields from the electrodes are not screened and extend toward each other through the solution. If a particle such as a cell passes through this field, it changes the permittivity of the medium carrying the electric field, and these events are detected as a capacitance change that is converted to a frequency signal and then to a voltage.

The techniques we have described so far work by out-pacing the dynamics of the double layer, but an alternative approach was shown in simulations by Liu et al[31,48] (Fig. 3D). In this work, a finite element method simulator was used to solve the PNP system for the case of a charged molecule passing through a nanopore. The authors showed that the presence of electro-diffusion current running tangentially to the detection interface allows the potential generated by the target to persist far beyond the predictions of the Poisson-Boltzmann model. Subsequent simulations showed that applying this technique to a nanowire biosensor can yield an order of magnitude improvement in sensitivity relative to a nanopore when detecting a 12-nucleotide DNA strand[31]. This technique has not been demonstrated experimentally, but it is appealing because it does not require high-frequency instrumentation but rather relies on a DC ionic electro-diffusion current.

**Outlook**

The capability to electronically measure biomolecules offers an exciting opportunity for medical diagnostics and analytical chemistry, with the potential to achieve rapid, sensitive, and quantitative measurements in a low-cost device. Electric double layer charge screening has posed a formidable impediment to progress in developing such sensors for real-world use, but the research community is beginning to gain mechanistic insights that will enable us to circumvent this barrier. In this article, we have presented two new heuristics for bioelectronic detection beyond the Debye length: Debye volume and non-equilibrium measurement. For Debye volume-based methods, the physical design of the electrode-electrolyte interfaces can be tuned to reduce the impact of screening. In parallel, other groups are applying external stimuli to disrupt electric double layers and then collect measurements under non-equilibrium conditions and thereby achieve detection. In fact, both approaches can be applied at the same time: non-equilibrium techniques place no requirements on the electrode-electrolyte interface and allow for the use of previously developed surface chemistries for immobilization and other Debye volume-oriented strategies. Importantly since the core physical principles underlying these approaches are now reasonably well understood, it should be feasible to predictably design electronic biosensors that can consistently achieve the high sensitivity required for molecular diagnostics and other point-of-care applications.

**Box 1: Origin of the Debye length and its limitations**

The Debye length, $\lambda_D$, is the characteristic distance for potential decay due to screening. This value is derived from the linearized Poisson-Boltzmann model, which describes screening as an equilibrium balance between drift and diffusion[6,7]. For a flat plane, this model is as follows:

$$\psi(x) = \psi_0 \exp\left(-\frac{x}{\lambda_D}\right) \text{ with } \lambda_D = \sqrt{\frac{\epsilon_0 \epsilon_r k_B T}{e^2 \sum z_i^2 n_{i\infty}}}$$

where $\psi_0$ is the potential at the interface, $x$ is the separation from the flat plane, $\epsilon_0$ and $\epsilon_r$ are the vacuum and relative permittivity, $k_B$ is the Boltzmann constant, $T$ is the temperature, $e$ is the elementary charge, and $z_i$ and $n_{i,\infty}$ are the valence number and bulk density of ion species $i$. To increase the Debye length, one can lower the electrolyte ionic strength. However, this comes at a cost, because diluting samples to lower the ionic strength also dilutes analyte concentrations, making the detection of low-abundance analytes more difficult.

It is important to recognize that the Debye length is derived from a simplified model of the double layer that does not fully capture the charge screening phenomenon. For example, $\lambda_D$ is only valid at that limit of small potentials where $z_i \psi \ll \frac{k_B T}{e}$ (the thermal voltage). This is called the Debye-Huckel approximation. These potentials are measured relative to the point of zero charge, a value that varies with electrode material and solution conditions. Another assumption inherent in $\lambda_D$ is that the system is at thermodynamic equilibrium – this allows the use of Boltzmann statistics to describe the density of ions. Furthermore, this derivation of $\lambda_D$ assumes that ions are point charges with negligible size but with finite diffusivity. This allows the model to discount any effects of ion crowding beyond diffusion. Although these conditions can be met under specific conditions, clearly $\lambda_D$ does not fully capture the charge screening phenomena under all circumstances.

To describe the capacitance created by the double layer, a parallel plate capacitor model is widely used with a separation of $\lambda_D$:

$$\sigma_s = \frac{\epsilon_0 \epsilon_r}{\lambda_D} \psi_0 \rightarrow C_{PB} = \frac{\epsilon_0 \epsilon_r}{\lambda_D} \frac{F}{m^2}$$

Although this is conceptually useful, there are important limitations to this description. This is because the double layer capacitance originates from two separate physical phenomena – the charges adsorbed to the electrode surface (the Stern layer) and the mobile ions in solution (the diffuse layer). Accordingly, double layer capacitance is actually highly non-linear, contrary to the simple model described above. Thus, although $\lambda_D$ is a useful parameter for conceptualizing charge screening in ionic solutions, it is also based on assumptions that do not hold under all conditions.

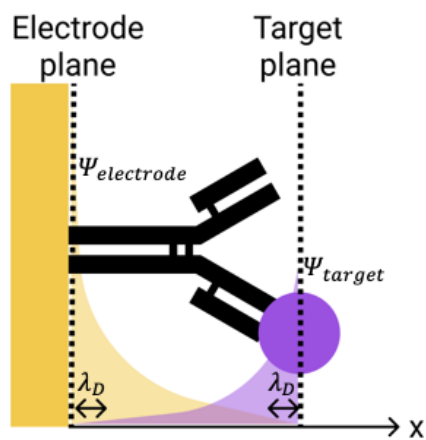

**Figure B1:** Electrical double layers form around all charges in electrolyte solutions. To generate an electronic signal, the target molecule's double layer (purple) must interact with the sensing electrode's double layer (yellow). If the target creates sufficiently high potentials, signals can be detected even when the target binds many Debye lengths from the electrode.

Finally, we note that using a capacitor to model the electric double layer, though useful for electronic interface design, implies that double layers form only in the vicinity of the electrode. Consequently, capacitive models describe target binding as a change in either surface charge ($\sigma_s$) or dielectric constant ($\epsilon_r$), but this cannot be the case if the target is far outside the electrode's double layer. But in fact, the signal generated by the analyte binding arises as the result of two interacting double layers—from both the electrode and the target (Fig. B1). For electrostatic biosensors, a more accurate explanation is that the screened electric fields of the adsorbed charges from the analyte are interacting with the surface, thereby changing the surface potential at the electrode interface. This effect also has important ramifications for capacitive sensors, as the adsorbed target and its double layer affect the configuration of the electrode's double layer, thereby altering the impedance.

**Acknowledgements**

We thank Evelin Sullivan and Michael Eisenstein for their thoughtful comments and edits on the manuscript.